\documentclass{article}
\usepackage[english]{babel}
\usepackage{amssymb}
\usepackage{amsfonts}
\usepackage{amsthm}
\usepackage{amsmath}
\usepackage{latexsym}

\usepackage{graphicx}
\usepackage{caption}
\usepackage{subcaption}
\usepackage[colorlinks=true,linkcolor=blue, citecolor=cyan, urlcolor=cyan]{hyperref}    
\usepackage{chngcntr}
\usepackage{caption}

\begin{document}

\title{{\Large Dividing goods \textbf{and} bads under additive utilities%
\thanks{%
Support from the Basic Research Program of the National Research University
Higher School of Economics is gratefully acknowledged. Sandomirskiy is
partially supported by the grant 16-01-00269 of the Russian Foundation for
Basic Research. }}}
\author{{\large Anna Bogomolnaia}$^{\bigstar \spadesuit }${\large , Herv\'{e}
Moulin}$^{\bigstar \spadesuit }${\large ,} \and {\large Fedor Sandomirskiy}$%
^{\spadesuit } ${\large , and Elena Yanovskaya}$^{\spadesuit }${\large .}
}
\date{{\large $^{\bigstar } $ \textit{University of Glasgow}\\
\vskip 0.1cm 
$^{\spadesuit }$ 
\textit{Higher School of Economics, St Petersburg}}}
\maketitle

\begin{abstract}
When utilities are additive, we uncovered in our previous paper \cite{BMSY}
many similarities but also surprising differences in the behavior of the
familiar \textit{Competitive rule }(with equal incomes), when we divide
(private) \textit{goods} or\textit{\ bads}.The rule picks in both cases the
critical points of the product of utilities (or disutilities) on the
efficiency frontier, but there is only one such point if we share goods,
while there can be exponentially many in the case of bads.

We extend this analysis to the fair division of \textit{mixed items}: each
item can be viewed by some participants as a good and by others as a bad,
with corresponding positive or negative marginal utilities. We find that the
division of mixed items boils down, normatively as well as computationally,
to a variant of an \textit{all goods} problem, or of an \textit{all bads}
problem: in particular the task of dividing the non disposable items must be
either good news for everyone, or bad news for everyone.

If at least one feasible utility profile is positive, the Competitive rule
picks the unique maximum of the product of (positive) utilities. If no
feasible utility profile is positive, this rule picks all critical points of
the product of \textit{dis}utilities on the efficient frontier.
\end{abstract}

\section{Introduction}

In our previous paper \cite{BMSY} we consider fair division of (private,
divisible) items under linear preferences, represented for convenience by
additive utilities. We explain there the appeal of this domain restriction
for the practical implementation of division rules vindicated by theoretical
analysis. We focus there on the \textit{Competitive Rule} (aka Competitive
Equilibrium with Equal Incomes) to divide the items, and contrast its
behavior when we divide \textit{goods }(assets, such as family heirlooms,
real estate, land, stocks), and when we divide \textit{bads }(chores,
workloads, liabilities, noxious substances or facilities). Several normative
properties of this rule are identical in both contexts, e. g., \textit{No
Envy} and a simple version of Maskin Monotonicity that we call \textit{%
Independence of Lost Bids}. However the unexpected finding is that several
aspects of the rule are very different in the two contexts: dividing \textit{%
bads} is not a mirror image of dividing \textit{goods.} The Competitive Rule
picks a unique welfare profile when it divides goods, but for dividing bads
it often proposes many (up to exponentially many in the number of agents and
bads) allocations with different welfare consequences; in the former case
the competitive welfare profile is continuous in the marginal rates of
substitution, in the latter case such continuity is not feasible. Also the
rule makes every participant benefit from an increase in the goods to
divide, a monotonicity property that is out of reach when we divide bads.

Here we generalize this analysis to fair division problems involving (non
disposable) \textit{mixed items, }i. e., both goods and bads, or even items
about which participants disagree whether they are \textit{good} or \textit{%
bad}. An inheritance may include good and bad real estate (e. g., heavily
mortgaged or not), divorcing couples must allocate jewellery as well as
obnoxious pets, workers sharing a multiple jobs relish certain jobs and
loath others; and managers facing the division of onerous tasks may
deliberately add some desirable items to \textquotedblleft
sweeten\textquotedblright\ the deal of the workers.

For a start we show that, upon adapting the standard definition to allow for
the coexistence of positive and negative prices and for individual budgets
of arbitrary sign, the Competitive Rule is always non empty, and its basic
normative properties (\textit{No Envy, Independence of Lost Bids}, \textit{%
Core from Equal Split}\footnote{%
Although in a slightly weaker sense, see Lemma 1.}) are preserved.\footnote{%
Note that existence follows (much) earlier results about competitive
equilibrium under satiated (not necessarily linear) preferences such as \cite%
{MC}. Our proof in our special domain is however simpler and constructive.}

The \textit{status quo ante} situation with nothing to divide delivers in
our model zero utility to each participant. If all items are good, any
feasible allocation brings (weakly) positive utilities, so that the arrival
of the \textquotedblleft manna\textquotedblright\ is good news for everyone;
similarly the task of dividing non disposable, undesirable items is a chore
for everyone bringing weakly negative utilities to all. With mixed items,
some good, some bad, allocations where some participants enjoy positive
utility and others negative utility are of course feasible, and\ some
interpretations of fairness pick such divisions (see an example below).
Remarkably the Competitive Rule never does: in \textit{any} problem mixing
goods and bads, \textit{either} it weakly improves the welfare of \textit{all%
} agents above the status quo ex ante, \textit{or} it weakly decrease
everyone's welfare below this benchmark. The rule enforces a strong
solidarity among agents: the task of dividing any bundle of non disposable
items is either unanimously good news or unanimously bad news.

\paragraph{The result}

We call a problem \textit{positive} if the zero utility profile is strictly
below the efficient utility frontier; \textit{negative} if it is strictly
above; and \textit{null} if it is on this frontier. In a positive problem
the Competitive Rule (CR for short) picks the unique allocation maximizing
the Nash product of utilities among positive profiles: it behaves \textit{as
if }there are only goods, selects a unique utility profile, and enjoys the
same regularity and monotonicity properties as in the all-goods case (within
the class of positive problems). In a negative problem the CR picks all the
critical points of the product of \textit{dis}utilities over the efficient
negative profiles: it behaves \textit{as if }there are only bads, in
particular it may pick many different utility profiles, and loses its
regularity and monotonicity properties. In a null problem the rule
implements the null utility profile.

Note that the familiar \textit{Fair Share} utility still sets a lower bound
on each agent competitive utility, but these utilities are no longer a
useful benchmark as they can be of different signs in the same problem (see
an example below).

\paragraph{Some simple examples}

\begin{figure}[h]
\centering
\includegraphics[width=0.45\linewidth, clip=true, trim = 4.0cm 6.8cm 4.0cm 6.5cm]{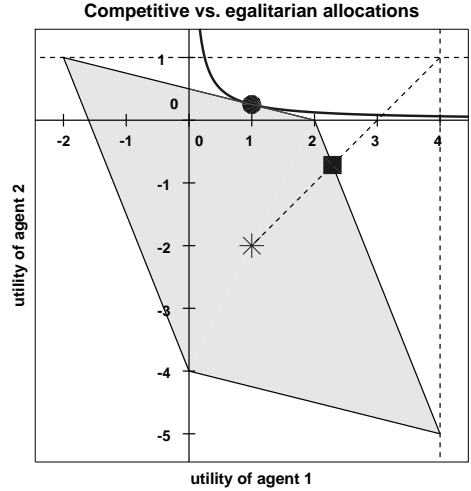}
\caption{Competitive (circle) and Egalitarian (square) utility profiles for the first example.}
\label{fig1}
\end{figure}

Here is a two-agent, two-item example where $a$ is a good and $b$ is a bad:%
\begin{equation*}
\begin{array}{ccc}
& a & b \\ 
u_{1} & 4 & -2 \\ 
u_{2} & 1 & -5%
\end{array}%
\end{equation*}%
There is one unit of each item to share. So $a$ is very good for agent $1$
compared to $b$, while $b$ is very bad relative to $a$ for agent $2$.%
\footnote{%
Recall interpersonal comparisons of utilities are ruled out, only the
underlying preferences matter.} Fair Share utilities obtain by giving one
half unit of each item to each agent: $U_{1}^{FS}=1,U_{2}^{FS}=-2$. The
familiar Egalitarian Rule equalizes the utility gains above this benchmark
relative to the maximal feasible gain:%
\begin{equation*}
\frac{U_{1}^{ER}-U_{1}^{FS}}{U_{1}^{MAX}-U_{1}^{FS}}=\frac{%
U_{2}^{ER}-U_{2}^{FS}}{U_{2}^{MAX}-U_{2}^{FS}}\text{ where }%
U_{1}^{MAX}=4,U_{2}^{MAX}=1
\end{equation*}%
Combined with Efficiency this gives $U_{1}^{ER}=2\frac{2}{7},U_{2}^{ER}=-%
\frac{5}{7}$: see Figure~\ref{fig1}. Thus the division task is good news for agent $1$
but not for agent $2$. By contrast the Competitive Rule focuses on the
interval of strictly positive and efficient utility profiles corresponding
to the allocations%
\begin{equation*}
\begin{array}{ccc}
& a & b \\ 
z_{1} & x & 1 \\ 
z_{2} & 1-x & 0%
\end{array}%
\text{ where }\frac{1}{2}\leq x\leq 1
\end{equation*}%
It picks the midpoint $x=\frac{3}{4}$ with corresponding utilities $%
U_{1}^{CR}=1,U_{2}^{CR}=\frac{1}{4}$, where agent $1$ gets only her Fair
Share utility.

\noindent Our next example is a negative problem%
\begin{equation*}
\begin{array}{ccc}
& a & b \\ 
u_{1} & 4 & -5 \\ 
u_{2} & 1 & -5%
\end{array}%
\end{equation*}%
where the efficient allocations with strictly negative utility profiles
cover the interval%
\begin{equation*}
\begin{array}{ccc}
& a & b \\ 
z_{1} & 0 & x \\ 
z_{2} & 1 & 1-x%
\end{array}%
\text{ where }\frac{4}{5}\leq x\leq 1
\end{equation*}%
so the competitive allocation is at $x=\frac{9}{10}$ with utilities $%
U_{1}^{CR}=U_{2}^{CR}=-\frac{1}{2}$, where again agent $1$ gets only her
Fair Share utility. The ER utilities are $U_{1}^{ER}=\frac{2}{5},U_{2}^{ER}=-%
\frac{7}{5}$.

\begin{figure}[h]
\centering

\includegraphics[width=0.6\linewidth, clip=true, trim = 2.5cm 7.9cm 2.5cm 7cm]{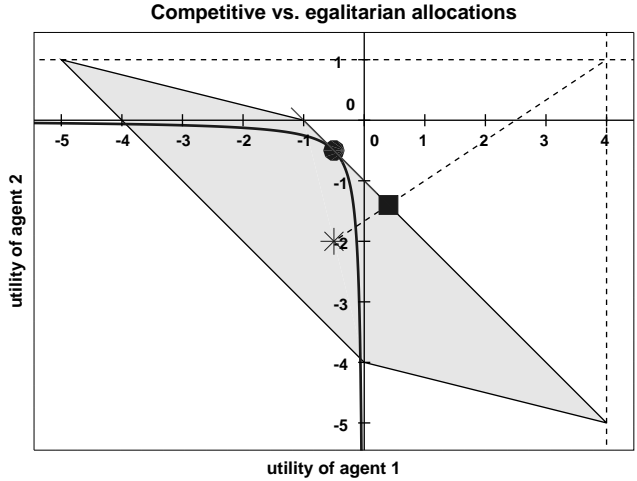}
\caption{Competitive (circle) and Egalitarian (square) utility profiles for the second example.}
\label{fig2}
\end{figure}

In Section 5 we compute the competitive division in a sequence of problems
with two agents, two bads, and a third item starting as a good and becoming
increasingly bad. The initial problem is positive, then becomes negative;
the number of competitive allocations takes all values from 1 to 4.

We stress another key difference between the Competitive and Egalitarian
Rules, implied by Independence of Lost Bids. If an object $a$ is a good for
some agents and a bad for others, efficiency implies that only the former
agents eat it. The CR ignores the detailed disutilities of the latter
agents: nothing changes if we set those disutilities to zero, so that $a$
becomes a good. This means that we need only to consider problems where
items are either (at least weakly) good for everyone, or (at least weakly)
bad for everyone. Obviously this simplification does not apply to the ER.

\section{The model}

The set of agents is $N$ with cardinality $n$, that of objects is $A$. A
problem is $\mathcal{P}=(N,A,u\in 
\mathbb{R}
^{N\times A})$ where the utility matrix $u$ has no null column.

With the notation $z_{M}=\sum_{i\in M}z_{i}$, and $e^{B}$ for the vector in $%
\mathbb{R}
^{B}$ with $e_{b}^{B}=1$ for all$b$, we define a feasible allocation as $%
z\in 
\mathbb{R}
_{+}^{N\times A}$ such that $z_{N}=e^{A}$. Let $\mathcal{F(P)}$ be the set
of feasible allocations, and $\Phi (\mathcal{P)}$ the corresponding set of
utility profiles. We always omit $\mathcal{P}$ if it is clear from the
context.

We call a feasible utility profile \textit{efficient }if it is not Pareto
dominated\footnote{%
That is $U\in \Phi (\mathcal{P)}$, and if $U\leq U^{\prime }$ and $U^{\prime
}\in \Phi (\mathcal{P)}$, then $U^{\prime }=U$.}; also a feasible allocation
is efficient if it implements an efficient utility profile.

The following two partitions, of $N$ and $A$ respectively, are critical:%
\begin{equation*}
N_{+}=\{i|\exists a:u_{ia}>0\},N_{-}=\{i|\forall a:u_{ia}\leq 0\}
\end{equation*}%
\begin{equation*}
A_{+}=\{a|\exists i:u_{ia}>0\},\text{ }A_{-}=\{a|\forall i:u_{ia}<0\},\text{ 
}A_{0}=\{a|\max_{i}u_{ia}=0\}
\end{equation*}%
When no confusion may arise, we call an object in $A_{+}$ \textit{good}, one
in $A_{-}$ \textit{bad}, and one in $A_{0}$ \textit{neutral.\smallskip }

\noindent \textbf{Definition 1:} \textit{For any problem }$\mathcal{P}$%
\textit{\ a competitive division is a triple }$(z\in \mathcal{F},p\in 
\mathbb{R}
^{A},\beta \in \{-1,0,+1\})$ \textit{where }$z$ \textit{is the competitive
allocation, }$p$\textit{\ is the price and }$\beta $\textit{\ the budget.
The allocation }$z_{i}$\textit{\ maximizes }$i$\textit{'s utility in the
budget set }$B(p,\beta )=\{y_{i}\in 
\mathbb{R}
_{+}^{A}|p\cdot y_{i}\leq \beta \}$:\textit{\ }%
\begin{equation}
z_{i}\in d_{i}(p,\beta )=\arg \max_{B(p,\beta )}\{u_{i}\cdot y_{i}\}
\label{3}
\end{equation}%
\textit{Moreover }$z_{i}$\textit{\ minimizes }$i$\textit{'s wealth in her
demand set}%
\begin{equation}
z_{i}\in \arg \min_{d_{i}(p,\beta )}\{p\cdot y_{i}\}  \label{12}
\end{equation}

\noindent \textit{The Competitive Rule selects at each problem} $\mathcal{P}$
\textit{the set} $CR(\mathcal{P)}$ \textit{of all competitive allocations}%
.\smallskip

In addition to the usual demand property (\ref{3}), we insist that an agent
spends as little as possible for her competitive allocation. This
requirement appears already in \cite{MC}: in its absence some satiated
agents in $N_{-}$ may inefficiently eat some objects useless to themselves
but useful to others. For instance in the two agents-two item problem $u=%
\begin{array}{cc}
6 & 2 \\ 
0 & -1%
\end{array}%
$ the inefficient allocation $z=%
\begin{array}{cc}
1/3 & 1 \\ 
2/3 & 0%
\end{array}%
$ meets (\ref{3}) for the prices\textit{\ }$p=(\frac{3}{2},\frac{1}{2})$ and
budget $\beta =1$. However $z_{2}=(0,0)$ guarantees the same (zero) utility
to agent $2$ and costs zero, so it fails (\ref{12}). The only competitive
division according to the Definition is $z=%
\begin{array}{cc}
1 & 1 \\ 
0 & 0%
\end{array}%
$ for $p=(\frac{1}{2},\frac{1}{2})$.

Check that in a competitive division we have%
\begin{equation}
p_{a}>0\text{ for }a\in A_{+}\text{ ; }p_{b}<0\text{ for }b\in A_{-}\text{ ; 
}p_{a}=0\text{ for }a\in A_{0}  \label{1}
\end{equation}%
If the first statement fails an agent who likes $a$ would demand an infinite
amount; if the second fails no one would demand $b$. If the third fails with 
$p_{a}>0$ the only agents who demand $a$ have $u_{ia}=0$, so that eating
some $a$ violates (\ref{12}); if it fails with $p_{a}<0$ an agent such that $%
u_{ia}=0$ gets an arbitrarily cheap demand by asking large amounts of $a$,
so (\ref{12}) fails again.

Here is another consequence of (\ref{12}), the importance of which is
illustrated by the above example:%
\begin{equation}
\forall a\in A_{+}:z_{ia}>0\Longrightarrow u_{ia}>0  \label{2}
\end{equation}%
Indeed if $i$ eats some $a\in A_{+}$ and $u_{ia}=0$, she gets a cheaper
competitive demand by ignoring $a$; and if $u_{ia}<0$ her allocation is not
competitive (recall $p_{a}>0$).\smallskip

We recall three standard normative properties of an allocation $z\in 
\mathcal{F(P)}$. It is \textit{Non Envious} iff $u_{i}\cdot z_{i}\geq
u_{i}\cdot z_{j}$ for all $i,j$; it guarantees the \textit{Fair Share}
utilities iff $u_{i}\cdot z_{i}\geq u_{i}\cdot (\frac{1}{n}e^{A})$ for all $%
i $. It is in the \textit{Weak Core from Equal Split} if for all $S\subseteq
N$ and all $y\in 
\mathbb{R}
_{+}^{S\times A}$ such that $y_{S}=\frac{|S|}{n}e^{A}$ there is at least one 
$i\in S$ such that $u_{i}\cdot z_{i}\geq u_{i}\cdot y_{i}$. When we divide
goods competitive allocations meet these three properties, even in the much
larger Arrow-Debreu preference domain.\smallskip

\noindent \textbf{Lemma 1 }\textit{A competitive allocation is efficient; it
meets No Envy, guarantees the Fair Share utilities, and is in the Weak Core
from Equal Split.\smallskip }

\noindent \textbf{Proof}.

\noindent \textit{Efficiency}. The classic argument by contradiction can be
adapted here. Let $z$ be a competitive allocation Pareto inferior to the
feasible allocation $y$. Some agent $i^{\ast }$ strictly prefers $y_{i^{\ast
}}$ to $z_{i^{\ast }}$ which implies $p\cdot z_{i^{\ast }}<p\cdot y_{i^{\ast
}}$ by (\ref{3}). So if we show $p\cdot z_{i}\leq p\cdot y_{i}$ for all $i$,
we contradict $z_{N}=y_{N}$ by summing up these inequalities. Note that we
can assume that $y$ itself is efficient which will be useful below.

First we have $p\cdot z_{i}=\beta $ for $i\in N_{+}$, or $i$ could buy more
of an object in $A_{+}$ he likes; moreover $i$ prefers strictly $z_{i}$ to
any $y_{i}$ such that $p\cdot y_{i}<\beta $, and weakly if $p\cdot y_{i}\leq
\beta $. So $p\cdot z_{i}\leq p\cdot y_{i}$ for all $i\in N_{+}$. It remains
to show $p\cdot z_{j}\leq p\cdot y_{j}$ for all $j\in N_{-}$.

We distinguish two cases. If $\beta =0,+1$ we have $u_{j}\cdot z_{j}=0$ (the
best feasible utility for $j$) and $u_{j}\cdot y_{j}=0$ as well. At $z$
agent $j$ can only consume objects in $A_{0}$: by (\ref{2}) $j$ eats no item
in $A_{+}$ and eating in $A_{-}$ strictly lowers her utility; agent $j$ eats
no object in $A_{+}$ at $y$ either by efficiency of $y$, and by $u_{j}\cdot
y_{j}=0$ she eats nothing in $A_{-}$ as well. Objects in $A_{0}$ are free ((%
\ref{1})) so $p\cdot z_{j}=p\cdot y_{j}=0$.

Now if $\beta =-1$ an agent $j$ in $N_{-}$ must eat some objects in $A_{-}$
that he dislikes hence his competitive demand $z_{j}$ has $p\cdot
z_{j}=\beta $ and as above he strictly prefers $z_{j}$ to $y_{j}$ if $p\cdot
y_{j}<\beta $: so $p\cdot z_{j}\leq p\cdot y_{j}$ as desired.

\noindent \textit{Other properties}. No Envy is clear and it implies Fair
Share by additivity of utilities. We use again the standard argument to
check the Weak Core property. Assume coalition $S\subset N$ has an objection 
$y$ to the competitive division $(z,p,\beta )$ where everybody in $S$
strictly benefits. So $y\in 
\mathbb{R}
_{+}^{S\times A}$ and $u_{i}\cdot z_{i}<u_{i}\cdot y_{i}$ for all $i\in S$.
If $\beta =0,+1$, this inequality is impossible for $i\in N_{-}$ because $%
u_{i}\cdot z_{i}=0$ (see above), so $S\subseteq N_{+}$. Then we sum over $S$
the inequalities $p\cdot z_{i}<p\cdot y_{i}$ to get%
\begin{equation*}
|S|\beta =p\cdot z_{S}<p\cdot y_{S}=p\cdot \frac{|S|}{n}e^{A}
\end{equation*}%
which contradicts $p\cdot e^{A}=p\cdot z_{N}\leq n\beta $.

If $\beta =-1$ we have $p\cdot z_{i}=\beta $ for all $i$, which simplifies
the argument.$\blacksquare \smallskip $

\textit{Remark 1:} For positive problems a competitive allocation may fail
the standard Core from Equal Split property, where coalition $S$ blocks
allocation $z$ if it can use its endowment $\frac{|S|}{n}e^{A}$ to make
everyone in $S$ weakly better off and at least one agent strictly more. This
is because \textquotedblleft equal split\textquotedblright\ gives resources
to the agents in $N_{-}$ that they have no use for. Say three agents share
one unit of item $a$ with $u_{1}=u_{2}=1,u_{3}=-1$. Here CR splits $a$
between agents $1$ and $2$, which coalition $\{1,3\}$ blocks by giving $%
\frac{2}{3}$ of $a$ to agent $1$.

\textit{Remark 2: }It is easy to check that CR meets \textit{Independence of
Lost Bids, }the translation of Maskin Monotonicity under linear preferences:
see the precise definition in \cite{BMSY}. Just as in Proposition 2 of that
paper, CR is characterized by, essentially, combining this property with
Efficiency.

\section{The result}

The key to classify our problems when $N$ and $A$ are given is the relative
position of the set of feasible utility profiles $\Phi $ and the cone $%
\Gamma (N)=%
\mathbb{R}
_{+}^{N_{+}}\times \{0\}^{N_{-}}$, which can only be of three types. Write
the relative interior of $\Gamma $ as $\Gamma ^{\ast }(N)=%
\mathbb{R}
_{++}^{N_{+}}\times \{0\}^{N_{-}}$.\smallskip

\noindent \textbf{Definition 2 }\textit{We call the problem }$\mathcal{P}%
=(N,A,u)$

\noindent \textit{positive if }$\Phi (\mathcal{P})\cap \Gamma ^{\ast
}(N)\neq \varnothing $;

\noindent \textit{negative if }$\Phi (\mathcal{P})\cap \Gamma
(N)=\varnothing $;

\noindent \textit{null if }$\Phi (\mathcal{P})\cap \Gamma (N)=\{0\}$%
.\smallskip

We fix $\mathcal{P}$ and check that these three cases are exhaustive and
mutually exclusive. This amounts to show that $\Phi \cap \Gamma ^{\ast
}=\varnothing $ and $\Phi \cap \partial \Gamma \neq \varnothing $ together
imply $\Phi \cap \Gamma =\{0\}$. Pick $U$ non zero in $\Phi \cap \partial
\Gamma $ and derive a contradiction. Let $U_{i}>0$ for the agents in $%
P\subset N_{+}$ and $U_{j}=0$ for those in $Q=N_{+}\diagdown P$. If some $%
i\in Q$ eats some $a$ he likes ($u_{ia}>0$), he must also eat some $b$ he
dislikes ($u_{ib}<0$): then let someone in $P$ take a small amount of $b$
from $i$ and we get a new $U^{\prime }\in \Phi \cap \partial \Gamma $ where $%
P^{\prime }$ is larger than $P$. \ If no $j$ in $Q$ eats any $a$ she likes
(so she does not eat any she dislikes either), we pick any $i\in Q$ and an
item $a$ she likes; $a$ must be eaten at $U$ exclusively by some agents in $%
P\cup N_{-}$; if some $j\in P$ eats $a$ we let $j$ give a small amount of $a$
to $i$ and we have found $U^{\prime }\in \Phi \cap \partial \Gamma $ with a
larger $P^{\prime }$; if some $k$ in $N_{-}$ eats some $a$ we have $u_{ka}=0$
so again $k$ can give his share of $a$ to $i$ and $P$ increases. Repeating
this construction until $P=N_{+}$ we reach $U\in \Phi \cap \Gamma ^{\ast }$,
the desired contradiction.\smallskip

Given a smooth function $f$ and a closed convex $C$ we say that $x\in C$%
\textit{\ is a critical point of }$f$\textit{\ in }$C$ if the supporting
hyperplane of the upper contour of $f$ at $x$ supports $C$ as well:%
\begin{equation*}
\forall y\in C:\partial f(x)\cdot y\leq \partial f(x)\cdot x\text{ or }%
\forall y\in C:\partial f(x)\cdot y\geq \partial f(x)\cdot x
\end{equation*}%
This holds in particular if $x$ maximizes or minimizes $f$ in $C$.\smallskip

In the statement we write $\Phi ^{eff}$ for the set of efficient utility
profiles, and $%
\mathbb{R}
_{=}^{N}$ for the interior of $%
\mathbb{R}
_{-}^{N}$.\smallskip

\textbf{Theorem }\textit{Fix a problem} $\mathcal{P}=(N,A,u)$.

\noindent $i)$\textit{\ The problem} $\mathcal{P}$ \textit{has a} \textit{%
competitive division with a positive budget if and only if it is positive.
In this case an allocation is competitive iff it maximizes the product }$%
{\Large \Pi }_{N_{+}}U_{i}$\textit{\ over }$\Phi \cap \Gamma ^{\ast }$%
\textit{; thus the corresponding utility profile is unique, positive in }$%
N_{+}$ \textit{and zero in }$N_{-}$\textit{.}

\noindent $ii)$\textit{\ The problem} $\mathcal{P}$ \textit{has a
competitive division with a negative budget if and only if it is negative.
In this case an allocation is competitive iff it is a critical point of the
product }$\Pi _{N}|U_{i}|$\textit{\ in }$\Phi $ \textit{that belongs to} $%
\Phi ^{eff}\cap \mathbb{R}_{=}^{N}$.\textit{\ All utilities are negative in
any competitive allocation.}

\noindent $iii)$\textit{\ The problem} $\mathcal{P}$ \textit{has a
competitive allocation with a zero budget if and only if it is null. In this
case} \textit{an allocation is competitive iff its utility profile is
zero.\smallskip }

Note that the Theorem implies in particular that $CR(\mathcal{P})$ is non
empty for all $\mathcal{P}$.

\section{Proof}

First we give a closed form description of the competitive demands per
Definition 1, i. e., the solutions of (\ref{3}) plus (\ref{12}).\smallskip 

\textbf{Lemma 2 }\textit{Fix} $\mathcal{P}$ \textit{a budget }$\beta \in
\{-1,0,+1\}$\textit{\ and a price }$p$\textit{\ such that }$p_{a}>0$\textit{%
\ in }$A_{+}$\textit{, }$p_{b}<0$\textit{\ in }$A_{-}$\textit{, and }$%
p_{a}=0 $\textit{\ in }$A_{0}$\textit{.}

\noindent $i)$ \textit{if }$i\in N_{-}$\textit{\ and }$\beta =0,+1$\textit{,
the allocation }$z_{i}\in 
\mathbb{R}
_{+}^{A}$\textit{\ is competitive iff }$u_{i}\cdot z_{i}=0$\textit{\ (for
instance }$z_{i}=0$\textit{)}

\noindent $ii)$\textit{\ if }$i\in N_{-}$\textit{\ and }$\beta =-1$\textit{, 
}$z_{i}\in 
\mathbb{R}
_{+}^{A}$\textit{\ is competitive iff }$p\cdot z_{i}=-1$\textit{, }$z_{ia}=0$%
\textit{\ on }$A_{+}$\textit{, }$z_{ia}>0$\textit{\ on }$A_{0}$\textit{\
only if }$u_{ia}=0$\textit{, and}%
\begin{equation}
\{b\in A_{-}\text{ and }z_{ib}>0\}\Longrightarrow \frac{|u_{ib}|}{|p_{b}|}%
\leq \frac{|u_{ib^{\prime }}|}{|p_{b^{\prime }}|}\text{ for all }b^{\prime
}\in A_{-}  \label{8}
\end{equation}%
$iii)$\textit{\ if }$i\in N_{+}$\textit{\ the problem (\ref{3}) has a
bounded solution iff}%
\begin{equation}
\frac{u_{ia}}{p_{a}}\leq \frac{|u_{ib}|}{|p_{b}|}\text{ for all }i\in
N_{+},a\in A_{+},b\in A_{-}  \label{10}
\end{equation}%
\textit{Then the allocation }$z_{i}$\textit{\ is competitive iff }$p\cdot
z_{i}=\beta $\textit{, and }$z_{i}$ \textit{meets (\ref{8}) and the two
following properties:}

\begin{equation}
\{a\in A_{+}\text{ and }z_{ia}>0\}\Longrightarrow \frac{u_{ia}}{p_{a}}\geq 
\frac{u_{ia^{\prime }}}{p_{a^{\prime }}}\text{ for all }a^{\prime }\in A_{+}
\label{7}
\end{equation}%
\begin{equation}
\{a\in A_{+},b\in A_{-},\text{ and }z_{ia}>0,z_{ib}>0\}\Longrightarrow \frac{%
u_{ia}}{p_{a}}=\frac{|u_{ib}|}{|p_{b}|}  \label{9}
\end{equation}

Statement $i)$ is clear upon noticing that eating some object in $A_{0}$ is
free so that (\ref{12}) holds. For $ii)$ observe that to meet the budget
constraint $i$ must be buying some $b\in A_{-}$; if $i$ buys some $a\in A_{+}
$ she can increase her utility by buying less of $b$ and of $a$; and (\ref%
{12}) still holds because objects in $A_{0}$ are free. Property (\ref{8})
simply says that she buys objects in $A_{-}$ with the smallest disutility
per unit of (fiat) money.

For statement $iii)$ pick agent $i$ in $N_{+}$ and note that a budget
balanced purchase of both objects $a\in A_{+}$ and $b\in A_{-}$ increases
strictly $i$'s utility iff $\frac{u_{ia}}{p_{a}}>\frac{|u_{ib}|}{|p_{b}|}$,
in which case (\ref{3}) has no bounded solution. As already noted in the
proof of Lemma 1, $i$ can buy some object he likes in $A_{+}$ with any slack
budget, therefore $p\cdot z_{i}=\beta $, which implies (\ref{12}).
Properties (\ref{8}) follow as for agents in $N_{-}$ and (\ref{7}) is
similar. Finally if $i$ eats both $a\in A_{+}$ and $b\in A_{-}$, inequality $%
\frac{u_{ia}}{p_{a}}<\frac{|u_{ib}|}{|p_{b}|}$ implies that a budget neutral
reduction of $z_{ia}$ and $z_{ib}$ increases $U_{i}$: thus we need (\ref{9})
as well. $\blacksquare \smallskip $

\textbf{Proof of the Theorem}

\noindent \textit{Step 1}: \textit{Statement }$\mathit{i)}$\textit{.} 
\textit{Let }$\mathcal{P}=(N,A,u)$\textit{\ be a positive problem}.\textit{\
We show that the maximization of the Nash product on }$\Phi \cap \Gamma $%
\textit{\ finds a Competitive allocation with a positive budget.}

\noindent As $\Phi \cap \Gamma $ is compact and convex, there is a unique $%
U^{\ast }$ maximizing in $\Phi \cap \Gamma $ the product ${\Large \Pi }%
_{N_{+}}U_{i}$; clearly $U_{i}^{\ast }>0$ for all $i\in N_{+}$. Let $z^{\ast
}$ be an allocation implementing $U^{\ast }$. By efficiency the items in $%
A_{0}$ are only eaten by agents in $N_{+}$ and/or $N_{-}$, who do not care
about them ($u_{ia}=0$); this implies%
\begin{equation}
\forall i\in N_{+}:u_{iA_{0}}\cdot z_{iA_{0}}^{\ast }=0  \label{4}
\end{equation}%
(with the notation $u_{iB}\cdot z_{iB}=\sum_{B}u_{ib}z_{ib}$).

By efficiency the items in $A_{+}$ are eaten in full by $N_{+}$ (property (%
\ref{2})); ditto for the items in $A_{-}$ because for such items and any $%
i\in N_{-}$, we have $u_{ia}<0$ and $U_{i}^{\ast }=0$. We define prices as
follows:%
\begin{equation}
\forall a\in A_{+}:p_{a}=\max_{N_{+}}\frac{u_{ia}}{U_{i}^{\ast }}>0;\text{ }%
\forall b\in A_{-}:p_{b}=-\min_{N_{+}}\frac{|u_{ib}|}{U_{i}^{\ast }}<0
\label{5}
\end{equation}%
and $p_{a}=0$ on $A_{0}$.

Pick any $a\in A_{+}$ and $i\in N_{+}$ eating $a$ (such $i$ exists by the
argument above): then $u_{ia}>0$ by efficiency, so the FOC of the
maximization program implies $\frac{\partial }{\partial z_{ja}}\ln
(u_{j}\cdot z_{j}^{\ast })\leq \frac{\partial }{\partial z_{ia}}\ln
(u_{i}\cdot z_{i}^{\ast })$ $\Longleftrightarrow \frac{u_{ia}}{U_{i}^{\ast }}%
\geq \frac{u_{ja}}{U_{j}^{\ast }}$ for all $j\in N_{+}$. This is property (%
\ref{7}). Checking (\ref{8}) is similar: assume $i\in N_{+}$ eats $b\in
A_{-} $ and recall $u_{ib}<0$, so the FOCs give $\frac{|u_{ib}|}{U_{i}^{\ast
}}\leq \frac{|u_{jb}|}{U_{j}^{\ast }}$ for all $j\in N_{+}$. Now we fix $%
i\in N_{+}$ and apply (\ref{7}), (\ref{8}):%
\begin{equation}
\{a\in A_{+}\text{ and }z_{ia}^{\ast }>0\}\Longrightarrow U_{i}^{\ast }=%
\frac{u_{ia}}{p_{a}}\text{ ; }\{b\in A_{-}\text{ and }z_{ib}^{\ast
}>0\}\Longrightarrow U_{i}^{\ast }=\frac{|u_{ib}|}{|p_{b}|}  \label{6}
\end{equation}%
Summing up numerator and denominator over the support of $z_{i}^{\ast }$
(the items he eats) and invoking (\ref{4}) as well as $p_{a}=0$ on $A_{0}$,
we get%
\begin{equation*}
U_{i}^{\ast }=\frac{\sum_{A_{+}}u_{ia}z_{ia}^{\ast
}-\sum_{A_{-}}|u_{ib}|z_{ib}^{\ast }}{\sum_{A}p_{a}z_{ia}^{\ast }}=\frac{%
u_{i}\cdot z_{i}^{\ast }}{p\cdot z_{i}^{\ast }}\Longrightarrow p\cdot
z_{i}^{\ast }=1
\end{equation*}%
\ as required by Lemma 2.

If $A_{-}$ is non empty, there is at least one agent $i\in N_{+}$ eating $%
a\in A_{+}$ and $b\in A_{-}$. For any such $i$ property (\ref{6}) gives $%
\frac{u_{ia}}{p_{a}}=U_{i}^{\ast }=\frac{|u_{ib}|}{|p_{b}|}$, which proves (%
\ref{9}), and 
\begin{equation*}
\text{for all }a^{\prime }\in A_{+},b^{\prime }\in A_{-}:\frac{u_{ia^{\prime
}}}{p_{a^{\prime }}}\leq U_{i}^{\ast }\leq \frac{|u_{ib^{\prime }}|}{%
|p_{b^{\prime }}|}
\end{equation*}%
implying (\ref{10}).\smallskip

\noindent \textit{Step 2: Statement }$\mathit{i)}$\textit{. Suppose the
problem} $\mathcal{P}=(N,A,u)$ \textit{has a competitive division} $(z,p,+1)$%
\textit{. We show that} $\mathcal{P}$ \textit{is positive and }$z$\textit{\
maximizes the Nash product as in Step 1.}

Because $z_{i}=0$ is in the budget set, all agents in $N_{-}$ must get zero
utility. If they consume anything, it must be an object in $A_{0}$ by
assumption (\ref{2}). Each $i$ in $N_{+}$ can buy some amount of any object,
so at $z$ her utility is positive: $U_{i}=u_{i}\cdot z_{i}>0$. Therefore $%
\mathcal{P}$ is positive.

Fix $i\in N_{+}$ and recall objects in $A_{0}$, if any, have zero price
(assumption (\ref{1})). Thus if $i$ eats some $a\in A_{0}$ we have $u_{ia}=0$%
, otherwise $i$ benefits by simply stop eating $a$. This gives $%
u_{iA_{0}}\cdot z_{iA_{0}}=0$ (as in (\ref{4})), and $p_{A_{0}}\cdot
z_{iA_{0}}=0$ as well.

We also know that $p$ is positive on $A_{+}$ and negative on $A_{-}$, and
that $p\cdot z_{i}=1$ for all $i\in N_{+}$ (by efficiency of $z$). Write the
sets of objects $i$ eats in $A_{+}\cup A_{-}$as $A_{+}(i)\cup A_{-}(i)$: $%
A_{+}(i)$ is non empty because $U_{i}>0$ ($A_{-}(i)$ can be empty). By Lemma
2 $\frac{u_{ia}}{p_{a}}$ is constant on $A_{+}(i)$, and equal to $\frac{%
|u_{ib}|}{|p_{b}|}$ on $A_{-}(i)$ if the latter is non empty. Thus this
common ratio is also%
\begin{equation}
\frac{u_{iA_{+}(i)}\cdot z_{iA_{+}(i)}+u_{iA_{-}(i)}\cdot z_{iA_{-}(i)}}{%
p_{A_{+}(i)}\cdot z_{iA_{+}(i)}+p_{A_{-}(i)}\cdot z_{iA_{-}(i)}}=\frac{U_{i}%
}{p\cdot z_{i}}=U_{i}  \label{11}
\end{equation}%
(where the first equality uses $u_{iA_{0}}\cdot z_{iA_{0}}=p_{A_{0}}\cdot
z_{iA_{0}}=0$). Therefore%
\begin{equation*}
\frac{u_{ia}}{U_{i}}=p_{a}\text{ for all }a\in A_{+}(i);\frac{|u_{ib}|}{U_{i}%
}=|p_{b}|\text{for all }b\in A_{-}(i)
\end{equation*}%
Then (\ref{7}) implies $\frac{u_{ia}}{p_{a}}\leq U_{i}$ for all $a\in A_{+}$
while (\ref{10}) implies $U_{i}\leq \frac{|u_{ib}|}{|p_{b}|}$ for all $b\in
A_{-}$. These two facts together give for all $i\in N_{+}$:%
\begin{equation}
\frac{u_{ia}}{U_{i}}\leq p_{a}\text{ for all }a\in A_{+}\text{ and }\frac{%
|u_{ib}|}{U_{i}}\geq |p_{b}|\text{ for all }b\in A_{-}  \label{15}
\end{equation}%
From this we derive that $U$ maximizes ${\Large \Pi }_{N_{+}}U_{i}$ in $\Phi
\cap \Gamma $, or equivalently that it is critical for the product of
utilities in $\Phi \cap \Gamma $: the restriction to $N_{+}$ of any feasible
utility profile is below the hyperplane supporting ${\Large \Pi }%
_{N_{+}}U_{i}$ at (the restriction of ) $U$:%
\begin{equation*}
\text{for all }U^{\prime }\in \Phi :\sum_{i\in N_{+}}\frac{U_{i}^{\prime }}{%
U_{i}}\leq n
\end{equation*}%
Pick $z^{\prime }\in \mathcal{F}$ implementing $U^{\prime }$ and use (\ref%
{15}) to compute (recalling that $p$ is zero on $A_{0}$)%
\begin{equation*}
\sum_{i\in N_{+}}\frac{u_{i}\cdot z_{i}^{\prime }}{U_{i}}=\sum_{a\in
A}\sum_{i\in N_{+}}\frac{u_{ia}z_{ia}^{\prime }}{U_{i}}\leq \sum_{a\in
A_{+}}\sum_{i\in N_{+}}p_{a}z_{ia}^{\prime }-\sum_{a\in A_{-}}\sum_{i\in
N_{+}}|p_{b}|z_{ib}^{\prime }=\sum_{a\in A}p_{a}=n
\end{equation*}

\noindent \textit{Step 3: Statement }$\mathit{ii)}$. \textit{Let }$\mathcal{P%
}=(N,A,u)$\textit{\ be a negative problem}.\textit{\ We show there exists a
critical point }$U^{\ast }$ \textit{of the product }$\Pi _{N}|U_{i}|$\textit{%
\ in }$\Phi $ \textit{that belongs to} $\Phi ^{eff}\cap \mathbb{R}_{=}^{N}$. 
\textit{The profile }$U^{\ast }$ \textit{we construct maximizes this product
on} $\Phi ^{eff}\cap \mathbb{R}_{-}^{N}$\textit{, and any allocation }$%
z^{\ast }$\textit{\ implementing it is competitive.}

\noindent \textit{Substep 3.1: If }$U^{\ast }\in \Phi ^{eff}\cap \mathbb{R}%
_{=}^{N}$ \textit{is a critical point of} $\Pi _{N}|U_{i}|$\textit{\ in }$%
\Phi $\textit{, then any }$z^{\ast }$\textit{\ implementing }$U^{\ast }$%
\textit{\ is competitive.}

We pick an allocation $z^{\ast }$ implementing $U^{\ast }$ and mimick the
argument in Step 1 above. While objects in $A_{+}$ are still eaten
exclusively by $N_{+}$, those in $A_{-}$ are eaten by anyone (and everyone
eats at least one object in $A_{-}$). If $a\in A_{+}$ and $z_{ia}^{\ast }>0$
for $i\in N_{+}$, a transfer of some $a$ from $i$ to $j\in N_{+}$ leaves the
allocation on the same side of $H$ as $\Phi $, i. e., below: this implies $%
\frac{u_{ia}}{|U_{i}^{\ast }|}\geq \frac{u_{ja}}{|U_{j}^{\ast }|}$; if $%
z_{ib}^{\ast }>0$ for $b\in A_{-}$ (and $i\in N$), we consider similarly a
transfer of some $b$ from $i$ to $j$ to get $\frac{|u_{ib}|}{|U_{i}^{\ast }|}%
\leq \frac{|u_{jb}|}{|U_{j}^{\ast }|}$. Then we define $p$ in $A_{+}\cup
A_{-}$ as in (\ref{5}), upon replacing $U^{\ast }$ by $|U^{\ast }|$ and
minimizing over all $N$ instead of just $N_{+}$ when defining $p$ in $A_{-}$. The analog of (\ref{6}) follows, with the same changes, and the same
computation yields $p\cdot z_{i}^{\ast }=-1$, this time for all $i$.

Setting $p_{a}=0$ on $A_{0}$, we now use Lemma 2 to check as in Step 1 that $%
z_{i}^{\ast }$ is $i$'s competitive for $p$ and $\beta =-1$.\smallskip 

\noindent \textit{Substep 3.2: We show that the profile }$U^{\ast }$\textit{%
\ maximizing} $\Pi _{N}|U_{i}|$\textit{\ in }$\Phi ^{eff}\cap \mathbb{R}%
_{-}^{N}$ \textit{is a critical point of this product\ in }$\Phi $ (and is
in $\mathbb{R}_{=}^{N}$).

We have $u_{i}\cdot e^{A}<0$ for every $i\in N$ , else the allocation $%
z_{i}=e^{A}$ yields utilities in $\Gamma $. Consider the set $F$ of utility
profiles dominated by $\Phi $: $F=\{U\in \mathbb{R}_{-}^{N}|\exists
U^{\prime }\in \Phi :U^{\prime }\leq U\}$. This set is closed and convex,
and contains all points in $\mathbb{R}_{-}^{N}$ that are sufficiently far
from the origin: any $U\in \mathbb{R}_{-}^{N}$ such that $U_{N}\leq
\min_{i}u_{i}\cdot e^{A}$ is dominated by the utility profile of $z:z_{i}=%
\frac{|U_{i}|}{|U_{N}|}e^{A},\ i\in N$.

Fix $\tau \geq 0$ and consider the upper contour of the Nash product at $%
\tau $: $K(\tau )=\{U\in \mathbb{R}_{-}^{N}|{\Large \Pi }_{N}|U_{i}|\geq
\tau \}$. For sufficiently large $\tau $ the closed convex set $K(\tau )$ is
contained in $F$. Let $\tau _{0}$ be the minimal $\tau $ with this property.
Negativity of $\mathcal{P}$ implies that $F$ is bounded away from $0$ so
that $\tau _{0}$ is strictly positive. By definition of $\tau _{0}$ the set $%
K(\tau _{0})$ touches the boundary of $F$ at some $U^{\ast }$ with strictly
negative coordinates. Let $H$ be a hyperplane supporting $F$ at $U^{\ast }$.
By the construction, this hyperplane also supports $K(\tau _{0})$, therefore 
$U^{\ast }$ is a critical point of the Nash product on $F$: that is, $%
U^{\ast }$ maximizes $\sum_{i\in N}\frac{U_{i}}{|U_{i}^{\ast }|}$ over all $%
U\in F$. So $U^{\ast }$ belongs to Pareto frontier of $F$, which is clearly
contained in the Pareto frontier of $\Phi $ Thus $U^{\ast }$ is a critical
point of the Nash product on $\Phi $ and belongs to $\Phi ^{eff}\cap \mathbb{%
R}_{=}^{N}$. Any $U$ in the interior of $K(\tau _{0})$ is clearly dominated
by some $U^{\prime }$ in $K(\tau _{0})\subset F$, hence by some $U^{\prime
\prime }\in \Phi \cap \mathbb{R}_{-}^{N}$: so $U^{\ast }$ maximizes the Nash
product on $\Phi ^{eff}\cap R_{-}^{N}$.

\textit{Remark 3:} Note that the supporting hyperplane $H$ to $\Phi $ at $%
U^{\ast }$ is unique because it is also a supporting hyperplane to $K(\tau
_{0})$ that is unique. This means that $U^{\ast }$ belongs to a face of a
polytope $\Phi $ of maximal dimension.\smallskip 

\noindent \textit{Step 4: Statement }$\mathit{ii)}$.\textit{\ Suppose the
problem} $\mathcal{P}=(N,A,u)$ \textit{has a competitive division} $(z,p,-1)$%
\textit{. We show that} $\mathcal{P}$ \textit{is negative and the
corresponding utility profile }$U$\textit{\ is a critical point of the
product }$\Pi _{N}|U_{i}|$\textit{\ in }$\Phi $ \textit{that belongs to} $%
\Phi ^{eff}\cap \mathbb{R}_{=}^{N}$.

The utility of any agent in $N_{-}$ at $z$ is negative: goods in $A_{0}$ are
free ((\ref{1})), he does not eat any object in $A_{+}$ ((\ref{2})), and his
budget is negative. Applying Lemma 2 to an agent in $N_{+}$ we see, as in
Step 2, that the ratio $\frac{u_{ia}}{p_{a}}$ is constant on $A_{+}(i)$, and
equal to $\frac{|u_{ib}|}{|p_{b}|}$ on $A_{-}(i)$ The same computation (\ref%
{11}) gives $\frac{u_{ia}}{p_{a}}=\frac{|u_{ib}|}{|p_{b}|}=\frac{U_{i}}{%
p\cdot z_{i}}=-U_{i}$, so $U_{i}<0$ in $N_{+}$ as well. By Lemma 1 the
negative profile $U$ is efficient, so $\Phi $ cannot intersect $\mathbb{%
\Gamma }$ and $\mathcal{P}$ is negative.

We derive now the criticality of $U$ for $\Pi _{N}|U_{i}|$ much like we did
in Step 2. The difference is that now \textit{everyone} eats some object in $%
A_{-}$, and $i$ in $N_{+}$ may or may not eat some object in $A_{+}$ (but $i$
in $N_{-}$ still doesn't).

From $|U_{i}|=\frac{u_{ia}}{p_{a}}$ on $A_{+}(i)$, $=\frac{|u_{ib}|}{|p_{b}|}
$ on $A_{-}(i)$ we get (\ref{12}) (for all $i$, and with $|U_{i}|$ instead
of $U_{i}$). Because any $i\in N$ must eat some $b^{\ast }$ in $A_{-}$,
properties (\ref{8}) and (\ref{12}) yield $|U_{i}|=\frac{|u_{ib^{\ast }}|}{%
|p_{b^{\ast }}|}\leq \frac{|u_{ib}|}{|p_{b}|}$ for all $b\in A_{-}$. Then (%
\ref{10}) gives $\frac{u_{ia}}{p_{a}}\leq |U_{i}|$ for all $a\in A_{+}$, and
this implies the analog of property (\ref{15}): for all $i\in N$%
\begin{equation}
\frac{u_{ia}}{|U_{i}|}\leq p_{a}\text{ for all }a\in A_{+}\text{ and }\frac{%
|u_{ib}|}{|U_{i}|}\geq |p_{b}|\text{ for all }b\in A_{-}  \label{16}
\end{equation}

The criticality of $U\in \Phi ^{eff}\cap \mathbb{R}_{=}^{N}$ for $\Pi
_{N}|U_{i}|$ in $\Phi $ means now that all feasible utility profiles are
below the hyperplane supporting ${\Large \Pi }_{N}|U_{i}|$ at $U$%
\begin{equation*}
\text{for all }U^{\prime }\in \Phi :\sum_{i\in N}\frac{U_{i}^{\prime }}{%
|U_{i}|}\leq -n
\end{equation*}%
(but this time $U$ does not maximize ${\Large \Pi }_{N}|U_{i}|$ on all of $%
\Phi $). The derivation of this inequality from (\ref{16}) proceeds exactly
as in Step 2.\smallskip 

\noindent \textit{Step 5: Statement }$iii)$\textit{. Let }$\mathcal{P}%
=(N,A,u)$\textit{\ be a null problem. We show there exists a price }$p$%
\textit{\ such that }$(z,p,0)$\textit{\ is competitive iff }$u_{i}\cdot
z_{i}=0$ for all $i$.

As $\Phi \cap \Gamma =\{0\}$ we can separate the projection of $\Phi $ on $%
\mathbb{R}
^{N_{+}}$ from $%
\mathbb{R}
_{+}^{N_{+}}$: there exists $\lambda \in 
\mathbb{R}
_{+}^{N_{+}}\diagdown \{0\}$ such that $\sum_{i\in N_{+}}\lambda
_{i}U_{i}\leq 0$ for all $U\in \Phi $. If $\lambda _{i}=0$ for some $i\in
N_{+}$ we pick $j\in N_{+}$ such that $\lambda _{j}>0$ and the allocation
where $j$ eats an object she likes and $i$ eats all the rest yields a
contradiction. Thus $\lambda $ is strictly positive.

Pick any $z^{\ast }\in \mathcal{F}$ implementing $U=0$: it is efficient
therefore $u_{iA_{0}}\cdot z_{iA_{0}}^{\ast }=0$. We have%
\begin{equation*}
z^{\ast }\in \arg \max \{\sum_{N_{+}}\lambda _{i}(u_{i}\cdot z_{i})|z\in 
\mathcal{F}\}=
\end{equation*}%
\begin{equation*}
=\arg \max {\Large \{}\sum_{a\in A_{+}}{\Large (}\sum_{N_{+}}\lambda
_{i}u_{ia}z_{ia}{\Large )}-\sum_{b\in A_{-}}{\Large (}\sum_{N_{+}}\lambda
_{i}|u_{ib}|z_{ib}{\Large )|}z\in \mathcal{F}{\Large \}}
\end{equation*}%
We define the price $p$ as%
\begin{equation*}
\forall a\in A_{+}:p_{a}=\max_{N_{+}}\lambda _{i}u_{ia};\text{ }\forall b\in
A_{-}:p_{b}=-\min_{N_{+}}\lambda _{i}|u_{ib}|
\end{equation*}%
and as usual $p=0$ in $A_{0}$. Clearly $p$ is positive on $A_{+}$and
negative on $A_{-}$. On the support of $z_{i}^{\ast }$ we have%
\begin{equation}
\forall a\in A_{+}\cup A_{-}:z_{ia}^{\ast }>0\Longrightarrow p_{a}=\lambda
_{i}u_{ia}  \label{17}
\end{equation}%
implying $p\cdot z_{i}^{\ast }=\lambda _{i}(u_{i}\cdot z_{i}^{\ast })=0$.
The definition of $p$ implies $p_{a}\geq \lambda _{i}u_{ia}$ and $%
|p_{b}|\leq \lambda _{i}|u_{ib}|$ for all $i\in N_{+},a\in A_{+},b\in A_{-}$%
. This implies (\ref{10}) at once; together with (\ref{17}) it gives (\ref{7}%
); the proof of (\ref{8}) and (\ref{9}) is similar.\smallskip 

\noindent \textit{Step 6: Statement }$\mathit{iii)}$.\textit{\ Suppose the
problem} $\mathcal{P}=(N,A,u)$ \textit{has a competitive division} $(z,p,0)$%
\textit{. We show that} $\mathcal{P}$ \textit{is null.}

Let $U$ be the utility profile of $z$. Clearly $U_{i}=0$ for $i\in N_{-}$.
Fix $i\in N_{+}$ such that $z_{i}\neq 0$: because $p\cdot z_{i}=0$ (Lemma 2) 
$i$ must eat at least one object in $A_{+}$ and one in $A_{-}$. By Lemma 2
again we have $\frac{u_{ia}}{p_{a}}=\frac{|u_{ib}|}{|p_{b}|}$ for all $a\in
A_{+}(i),b\in A_{-}(i)$. Writing $\frac{1}{\lambda _{i}}$ for this common
ratio, we have%
\begin{equation*}
u_{i}\cdot z_{i}=\frac{1}{\lambda _{i}}(p\cdot z_{i})=0
\end{equation*}%
and we conclude $U=0$. As $U\in \Phi ^{eff}$ the intersection $\Phi \cap
\Gamma $ contains nothing more.

\pagebreak

\section*{Appendix: a monotonic sequence of examples}

We have $N=\{1,2\}$, $A=\{a,b,c\}$ and%
\begin{equation*}
\begin{array}{cccc}
& a & b & c \\ 
u_{1} & -1 & -3 & \lambda  \\ 
u_{2} & -2 & -1 & \lambda 
\end{array}%
\end{equation*}%
and $\lambda $ takes all integer values from $4$ to $-3$. The first two
problems, for $\lambda =4,3$ are positive; for $\lambda =2$ the problem is
null, then negative from $\lambda =1$ to $-3$.

\vskip 0.9cm

\begin{figure}[h]
\centering
\begin{subfigure}[b]{0.48\linewidth}
\includegraphics[width=1\linewidth, clip=true, trim = 2.5cm 6.8cm 2.5cm 6.5cm]
{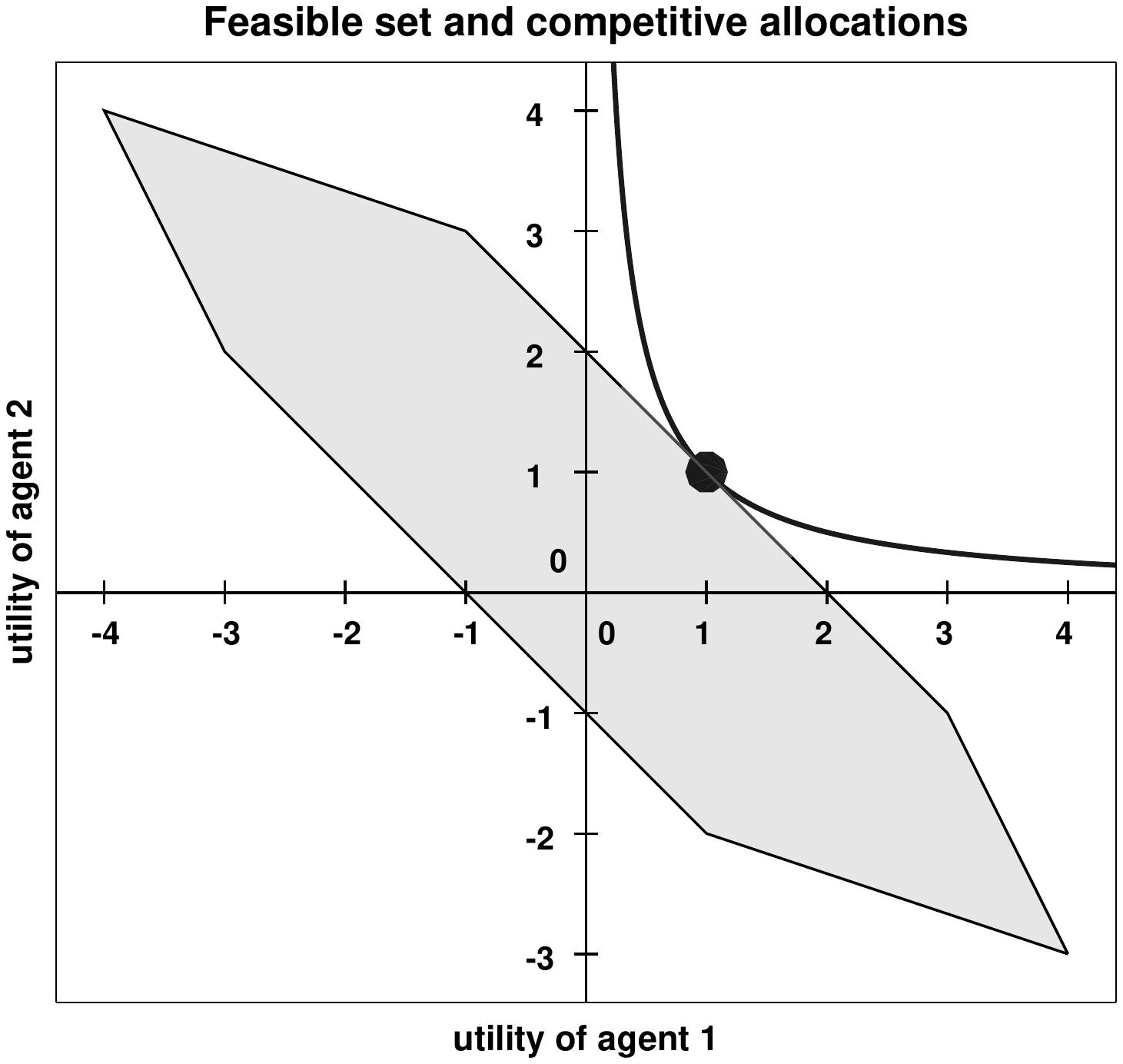}
\end{subfigure}
~
\begin{subfigure}[b]{0.48\linewidth}
\includegraphics[width=1\linewidth, clip=true, trim = 2.5cm 6.8cm 2.5cm 6.5cm]
{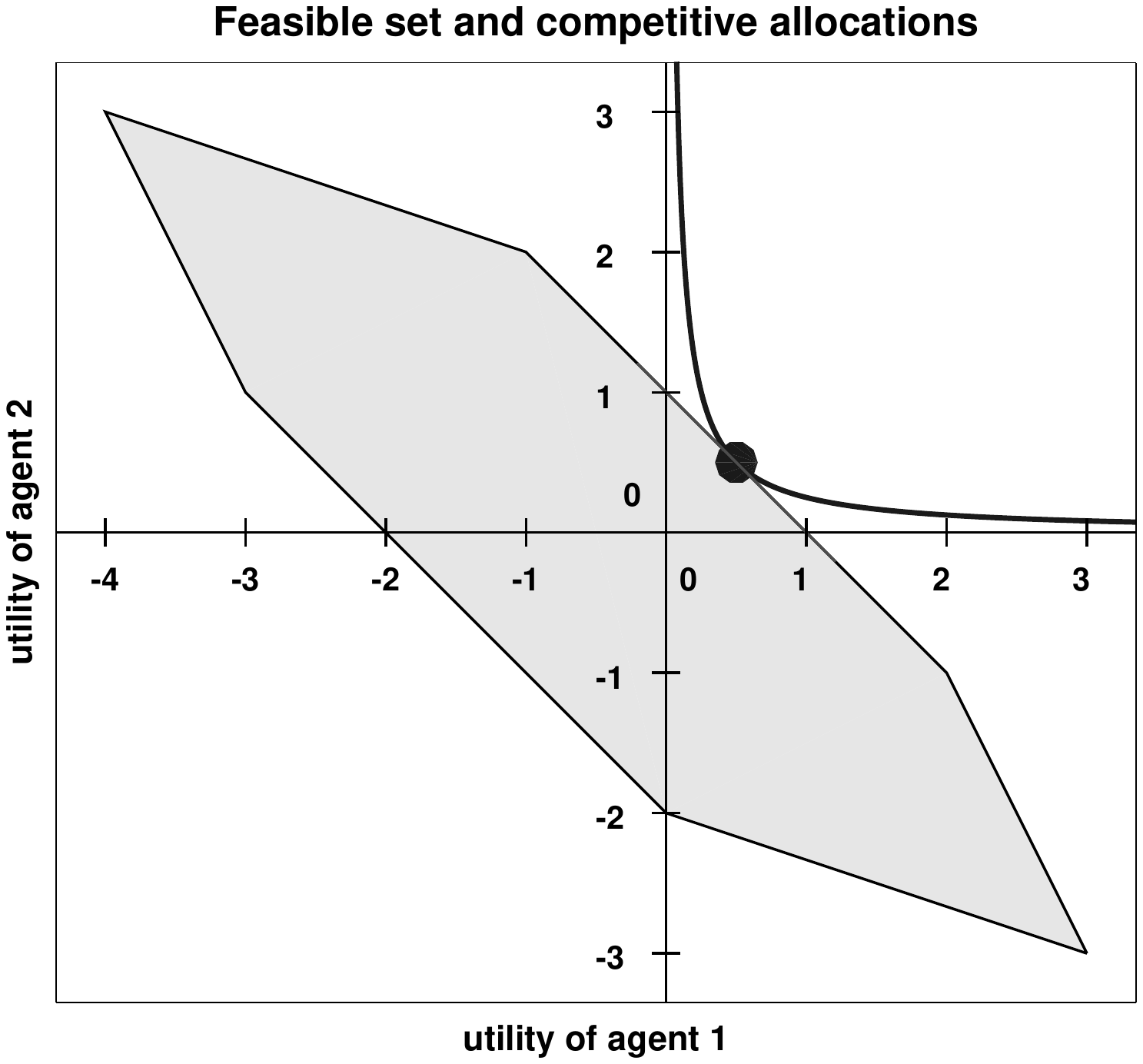}
\end{subfigure}
\\
\vskip 0.5cm
\begin{subfigure}[b]{0.48\linewidth}
\includegraphics[width=1\linewidth, clip=true, trim = 2.5cm 6.8cm 2.5cm 6.5cm]
{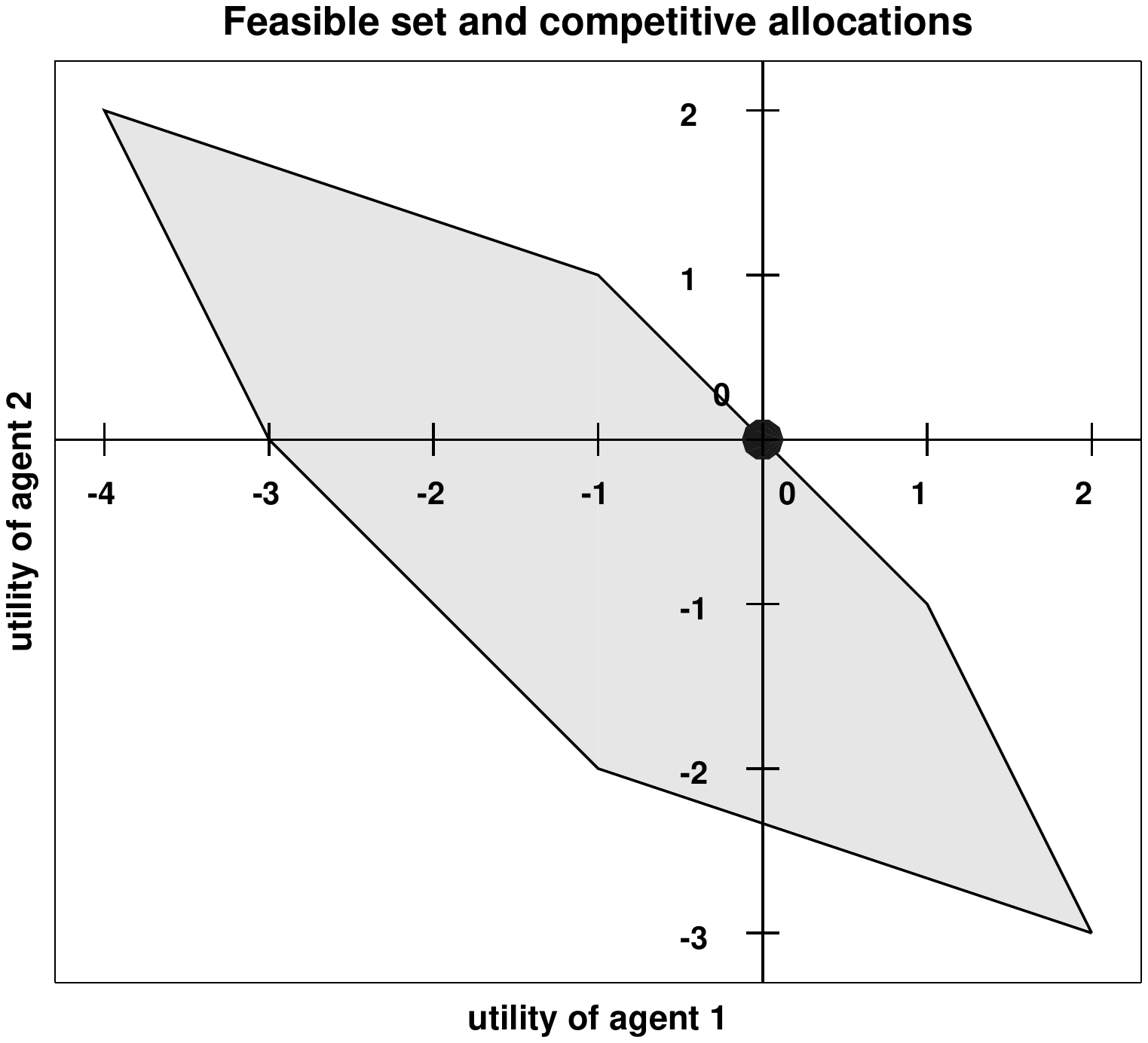}
\end{subfigure}
~
\begin{subfigure}[b]{0.48\linewidth}
\includegraphics[width=1\linewidth, clip=true, trim = 2.5cm 6.8cm 2.5cm 6.5cm]
{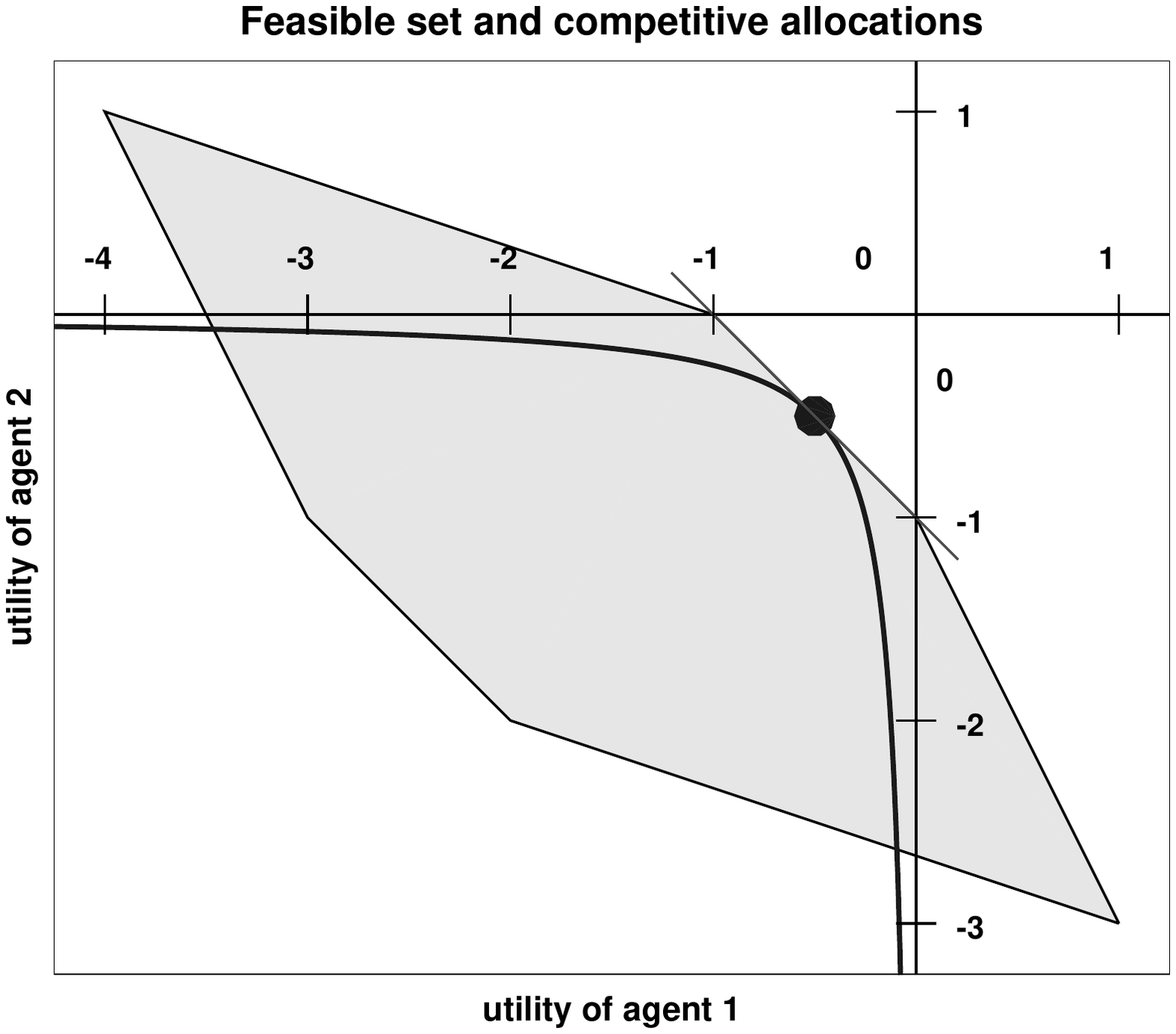}
\end{subfigure}
\end{figure}

\begin{figure}[h]
\centering
\begin{subfigure}[b]{0.48\linewidth}
\includegraphics[width=1\linewidth, clip=true, trim = 2.5cm 6.8cm 2.5cm 6.5cm]
{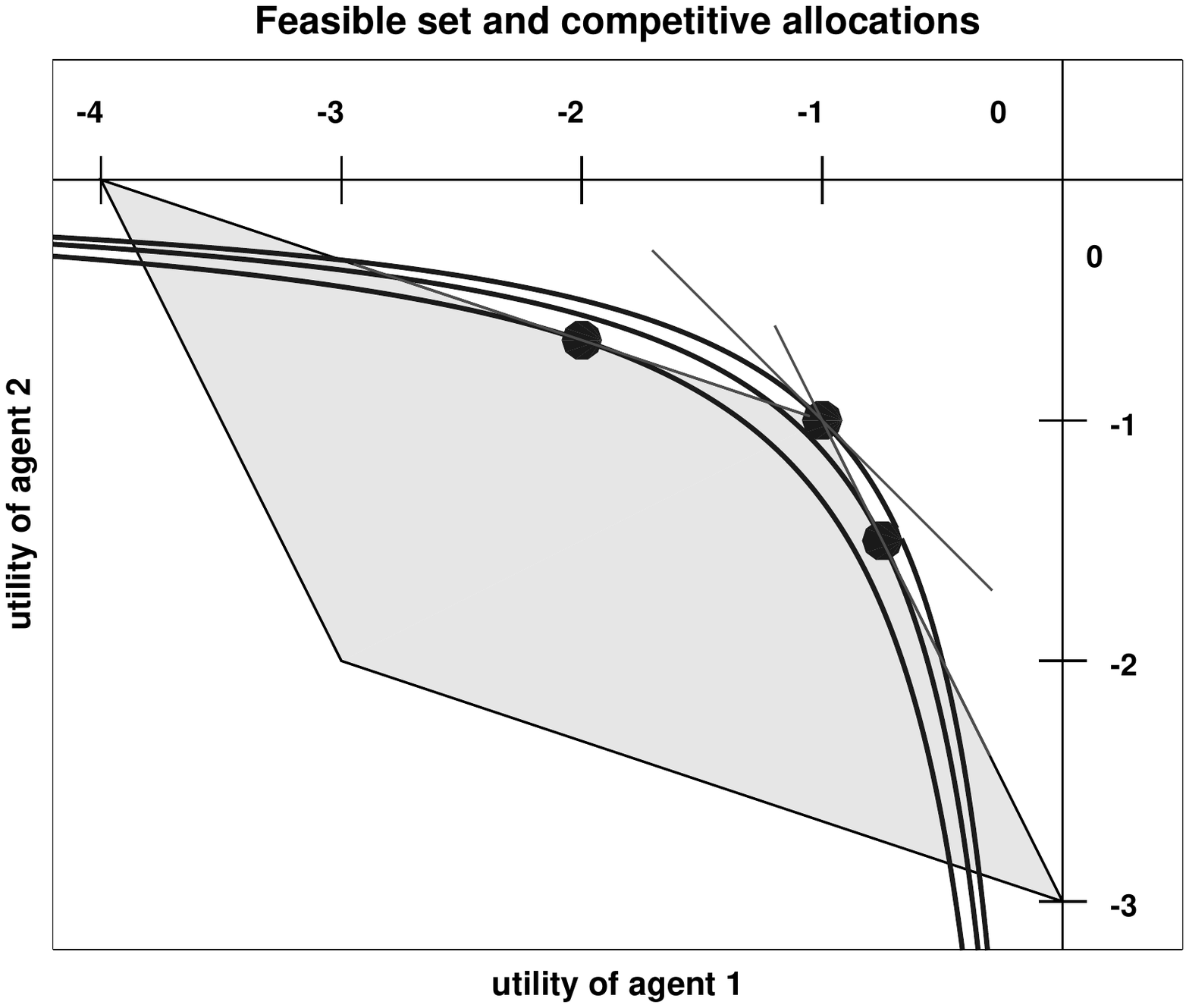}
\end{subfigure}
~
\begin{subfigure}[b]{0.48\linewidth}
\includegraphics[width=1\linewidth, clip=true, trim = 2.5cm 6.8cm 2.5cm 6.5cm]
{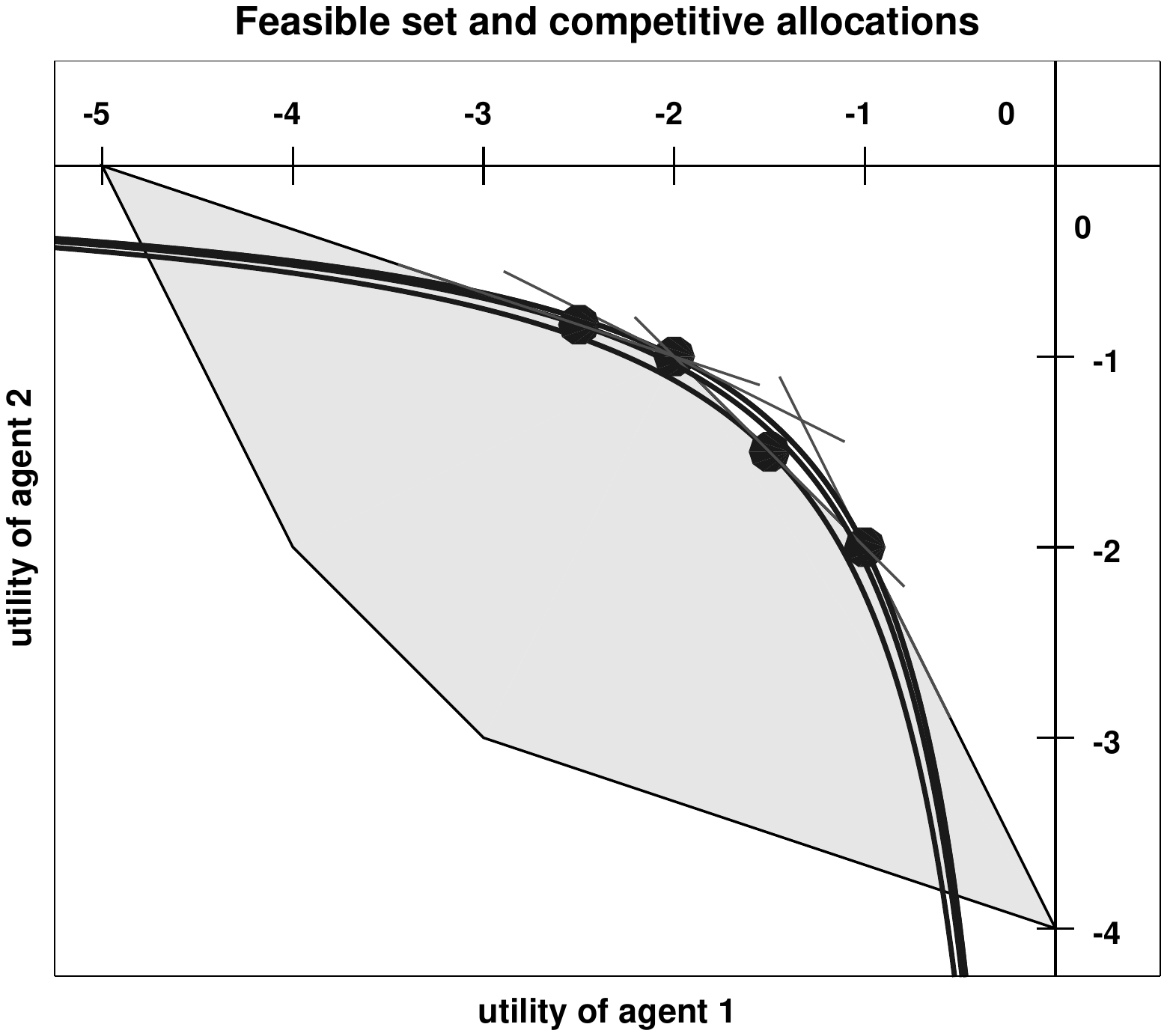}
\end{subfigure}
\\
\vskip 0.5cm
\begin{subfigure}[b]{0.48\linewidth}
\includegraphics[width=1\linewidth, clip=true, trim = 2.5cm 6.8cm 2.5cm 6.5cm]
{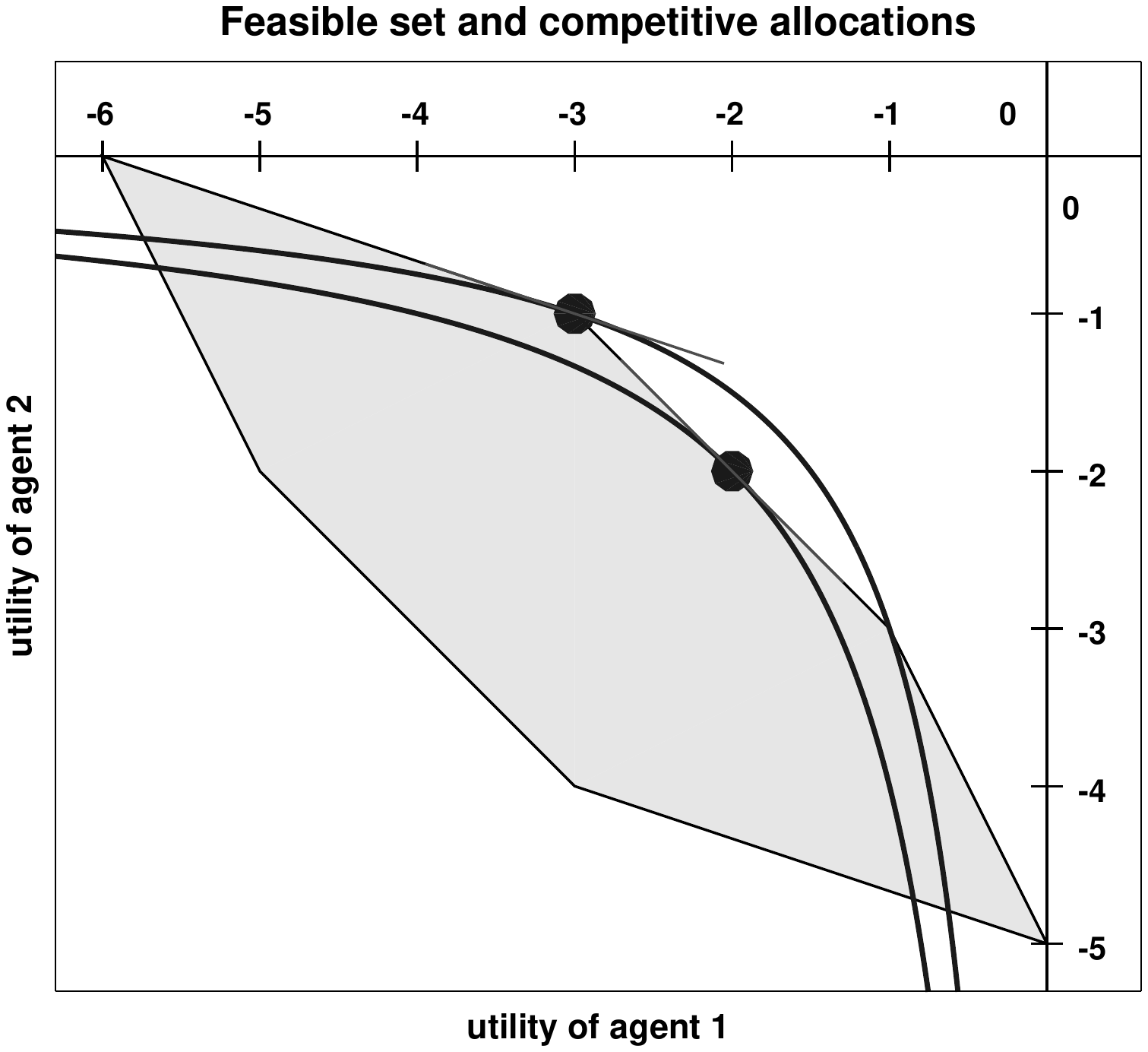}
\end{subfigure}
~
\begin{subfigure}[b]{0.48\linewidth}
\includegraphics[width=1\linewidth, clip=true, trim = 2.5cm 6.8cm 2.5cm 6.5cm]
{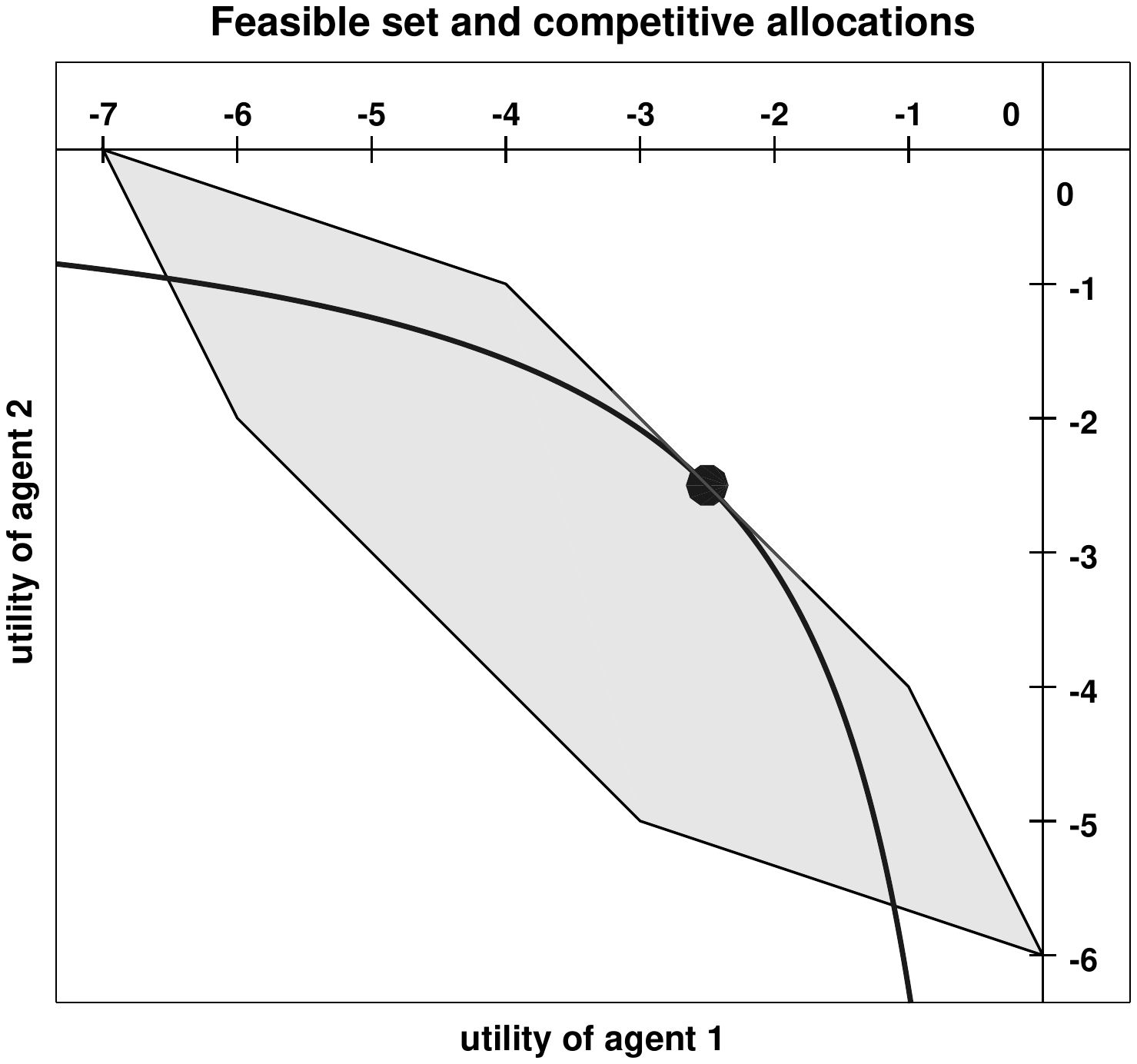}
\end{subfigure}

\end{figure}

\end{document}